\makeatletter
\def\input@path{{tex/}{bst/}}
\makeatother

\documentclass[twocolumn,superscriptaddress,amsmath,amssymb,aps,prb]{revtex4-2}

\usepackage[colorlinks=true,urlcolor=blue,citecolor=blue,linkcolor=blue]{hyperref}
\usepackage{txfonts}
\usepackage{graphicx}
\usepackage{bm}
\usepackage{bbm}
\usepackage{scalerel}
\usepackage{textcomp}

\usepackage{braket}
\usepackage{tikz}
\usetikzlibrary{shapes.geometric} 
\usepackage[mode=buildnew]{standalone}
\usetikzlibrary{backgrounds}

\renewcommand{\vec}[1]{\bm{#1}}
\newcommand{\ud}{\mathrm{d}}

\newcommand{\Eq}[1]{Eq.~(\ref{#1})}
\newcommand{\Fig}[1]{Fig.~\ref{#1}}

\begin{document}

\title{Automatic differentiation of dominant eigensolver \\ and its applications in quantum physics}

\author{Hao Xie}
\affiliation{Institute of Physics, Chinese Academy of Sciences, Beijing 100190, China}
\affiliation{University of Chinese Academy of Sciences, Beijing 100049, China}

\author{Jin-Guo Liu}
\affiliation{Institute of Physics, Chinese Academy of Sciences, Beijing 100190, China}

\author{Lei Wang}
\email{wanglei@iphy.ac.cn}
\affiliation{Institute of Physics, Chinese Academy of Sciences, Beijing 100190, China}
\affiliation{Songshan Lake Materials Laboratory, Dongguan, Guangdong 523808, China}

\date{\today}

\begin{abstract}
We investigate the automatic differentiation of dominant eigensolver where only a small proportion of eigenvalues and corresponding eigenvectors are obtained. Backpropagation through the dominant eigensolver involves solving certain low-rank linear systems without direct access to the full spectrum of the problem. Furthermore, the backward pass can be conveniently differentiated again, which implies that in principle one can obtain arbitrarily higher order derivatives of the dominant eigen-decomposition process. These results allow for the construction of an efficient dominant eigensolver primitive, which has wide applications in quantum physics. As a demonstration, we compute second order derivative of the ground state energy and fidelity susceptibility of 1D transverse field Ising model through the exact diagonalization approach. We also calculate the ground state energy of the same model in the thermodynamic limit by performing gradient-based optimization of uniform matrix product states. By programming these computational tasks in a fully differentiable way, one can efficiently handle the dominant eigen-decomposition of very large matrices while still sharing various advantages of differentiable programming paradigm, notably the generic nature of the implementation and free of tedious human efforts of deriving gradients analytically.
\end{abstract}

\maketitle

\section{\label{sec: Introduction}Introduction}

Automatic differentiation (AD) is a technique of numerically evaluating the exact derivatives of a computation process expressed as a program \cite{BARTHOLOMEWBIGGS2000171}. Basically, the derivatives are obtained by traversing through the computation graph from end to end and iteratively applying the chain rule. Compared to numerical differentiation, it can compute the value of derivatives to machine precision. Automatic differentiation is the computation engine underlying modern deep learning applications \cite{maclaurin2015gradient, rumelhart1986learning, baydin2018automatic} and has been realized in different ways in a large variety of modern deep learning libraries, such as TensorFlow \cite{tensorflow}, autograd \cite{autograd}, PyTorch \cite{pytorch}, Jax \cite{jax} and Zygote \cite{zygote}. This fact has triggered applications of AD in much broader research areas, such as quantum optimal control \cite{jirari2009optimal, PhysRevA.95.042318}, various electron structure methods in quantum chemistry \cite{doi:10.1021/acscentsci.7b00586, doi:10.1021/ct100117s, steiger2005using} and tensor network approach of studying statistical physics and quantum many-body problems \cite{PhysRevX.9.031041}.

One of the most important characteristics of automatic differentiation is its modular nature. More specifically, this means that the programmer can control the granularity of a computation process as he wants by grouping many ``elementary'' computation steps in a single unit. Such a unit often has higher-level mathematical meanings and is called a \textit{primitive}. For example, it turns out that many subroutines in scientific computing can be differentiated as a whole unit. Typical examples include solving ordinary differential equations \cite{chen2018neural}, various linear algebra manipulations, including both simple operations such as matrix multiplication, inverse and more sophisticated ones such as full eigen-decomposition, singular value decomposition (SVD), QR decomposition, etc \cite{giles2008extended, Seeger2017AutoDifferentiatingLA}. In view of this perspective, a new programming paradigm called \emph{differentiable programming} has emerged, which lays emphasize on assembling relative simple differentiable components (i.e., primitives) together and differentiating through them by applying the chain rule iteratively. By formulating a computational task in this way, one is able to combine domain-specific knowledge and the flexibility of modern machine learning techniques.

There can be two kinds of schemes of performing AD of a computation process, namely the forward mode and the reverse mode. The difference lies on the order of evaluating and passing gradients through the computation graph using the chain rule. The forward mode AD computes the gradients along with the objective output in a single forward pass, whereas in the reverse mode version, one needs an extra backward pass in which the gradient message are passed from the output back to the input via a series of vector-Jacobian products. This approach is usually referred to as the back-propagation algorithm \cite{rumelhart1986learning} in the context of deep learning. It is also favored and more commonly adopted in applications of physics and deep learning than the forward mode, due to the fact that the dimension of the output is often much smaller than input. For the same reason, we will almost exclusively focus on reverse mode AD in this paper.

Since primitives are the building blocks of a differentiable program, the central task of differentiable programming lies on AD of the primitives. To be more specific, in a typical computation process, one starts from some input parameters, say, $\theta$, and computes a series of intermediate results following the topological order characterized by the computation graph, until reaching the final outcome $\mathcal{L}$, which is usually assumed to be a scalar valued loss function. Consider a certain primitive generally described by a function $O = O(I)$, where $O$ and $I$ denote the outputs and inputs, respectively. It is convenient to introduce the \emph{adjoint} of a certain variable $T$ as $\overline{T} \equiv \frac{\partial \mathcal{L}}{\partial T}$. Then in reverse mode AD, one is typically concerned with the adjoint $\overline{I}$ of inputs as a function of the adjoint $\overline{O}$ of outputs. This is sometimes referred to as the \emph{adjoint relation} of the primitive and can be written in an abstract mathematical form as follows:
\begin{equation}
        \overline{I} \equiv \overline{I}(\overline{O}; I, O) = \overline{O} \frac{\partial O}{\partial I}.
\end{equation}
Notice that this function depends linearly on $\overline{O}$, as indicated by the linear approximation nature of derivatives. Once this function is determined for all the primitives involved in a computation process, one can then use the chain rule to ``glue'' them together and compute the desired gradient $\frac{\partial \mathcal{L}}{\partial \theta}$ by traversing through the computation graph in the reverse direction.

In this paper, we will concentrate on the automatic differentiation of dominant eigensolver, which is essentially the process of eigen-decomposition except that only a small number of eigenvalues and eigenvectors are desired. Eigen-decomposition plays a fundamental role in quantum physics and chemistry, and is related to many practical methods such as exact diagonalization, full configuration interaction and Hartree-Fock method \cite{doi:10.1021/acscentsci.7b00586}; it has also been used in conjunction with neural network architectures in various deep learning algorithms \cite{huang2018decorrelated, Ionescu2015, Dang2020, Zanfir_2018_CVPR, DBLP:journals/corr/abs-1906-09023}; furthermore, it has close and intrinsic relation with SVD, and they have been widely used in various tensor network algorithms \cite{ORUS2014117, SCHOLLWOCK201196, hasik2019towards, PhysRevX.9.031041}. In many of these applications, usually only a small number of eigenvalues and eigenvectors are of practical interest, despite the fact that the matrix dimension is possibly quite large. In the context of quantum physics and chemistry, for example, this typically means that one is concerned only about the ground state or several low-lying excited states. The situation is also similar in many other settings, including deep learning applications \cite{Zanfir_2018_CVPR, DBLP:journals/corr/abs-1906-09023} and tensor network algorithms \cite{ORUS2014117, SCHOLLWOCK201196}. In this respect, since the computational cost of full eigen-decomposition becomes fairly high when the matrix dimension is large, one would usually resort to more efficient numerical algorithms of dominant eigen-decomposition, such as power iteration or Lanczos method. These algorithms are particularly useful when the matrix to be diagonalized has certain inner structures (e.g., sparse), which is often the case in practical applications.

However, there emerges an additional difficulty when trying to implement the dominant eigen-decomposition process in a differentiable way. To get an intuitive understanding, consider a quantum system described by a Hamiltonian $H$, which depends on a certain parameter $\theta$. To obtain the derivative of the ground state $\Ket{\psi_0}$ with respect to $\theta$, the first order perturbation theory gives
\begin{equation}
    \Ket{\frac{\partial \psi_0}{\partial \theta}} = \sum_{n \neq 0} \frac{\Braket{\psi_n | \frac{\partial H}{\partial \theta} | \psi_0}}{E_0 - E_n} \Ket{\psi_n}.
    \label{eq: 1st order correction of ground state}
\end{equation}
where $E_n$, $\Ket{\psi_n}$ are the energy eigenvalues and eigenstates, respectively. Due to explicit presence of the full eigen-spectrum in \Eq{eq: 1st order correction of ground state}, the computation of full eigen-decomposition is inevitable in this direct approach, which is inefficient and even intractable when the dimension of the Hilbert space is large. One way around this problem is proposed in Ref.~\cite{DBLP:journals/corr/abs-1906-09023}, which is based on the power iteration algorithm of dominant eigensolver. Since the operations involved in the power iteration procedure are very simple (mainly matrix multiplications), it can be easily differentiated without any reference to the full spectrum. However, this approach also has some drawbacks. The convergence rate of power iteration can be a problem in practice and has to be analyzed case by case. In addition, although possible in principle, it is often tedious and impractical to obtain good estimate of other eigenvalues and eigenvectors than the dominant one through power iteration, which makes the approach inflexible to various user needs.

Basically, we need a way to effectively separate the information about the desired eigenvalues and eigenvectors out of the full spectrum. In this paper, this problem is tackled by two different methods, which allow us to construct a high-level primitive that correctly handles the AD of dominant eigensolver without direct access to the full spectrum. The first one, called the adjoint method, can yield the relevant formulas straightforwardly in a full-spectrum-free form. On the other hand, the second method reflects the modular nature of differentiable programming paradigm by wrapping the process of full eigen-decomposition within the dominant one and utilizing the results of the former in the latter. Typically, the results obtained by this method still have explicit dependence on the full spectrum. Nevertheless, these two methods are totally equivalent, and by making a careful contrast between them, one can get a clear understanding of how the goal of separating the desired information out of the full spectrum is achieved behind the scene. Even more ideally, it turns out that the obtained dominant eigensolver primitive could be differentiated again in a convenient way, which in turn makes it support in principle arbitrarily higher order derivatives of the dominant eigen-decomposition process. These results are very useful in practice, since they enable one to share the efficiency of dominant eigen-decomposition algorithms and various advantages of the differentiable programming paradigm discussed above at the same time.

The organization of this paper is as follows. In Sec. \ref{sec: Formulations}, the automatic differentiation of dominant eigensolver is studied in a typical setting. The mechanisms that effectively separate the desired information out of the full spectrum as well as support taking arbitrarily order of derivatives are carefully explained. In Sec. \ref{sec: Applications}, we demonstrate applications of the techniques by studying the ground state properties of 1D transverse field Ising model via two different approaches, namely exact diagonalization and gradient-based optimization of uniform matrix product states. The concluding remarks are given in Sec. \ref{sec: Discussions}. Our code implementation is publicly available~\cite{github}.

\section{\label{sec: Formulations}Formulations}

For the sake of simplicity and clarity, let $A$ be an $N$-dimensional real square matrix, and we are concerned with only one certain eigenvalue $\lambda$ and corresponding left and right eigenvector $\vec{l}$, $\vec{r}$ of $A$, respectively. In other words, we have
\begin{equation}
    \vec{l}^T A = \lambda \vec{l}^T, \quad A \vec{r} = \lambda \vec{r}, \quad \vec{l}^T \vec{r} = 1.
    \label{eq: relation between A, l, r}
\end{equation}
where we have imposed the conventional normalization condition. Note here that $A$ is generally non-symmetric, and we only assume that $A$ is diagonalizable and the desired eigenvalue $\lambda$ is non-degenerate. Since the set of non-diagonalizable matrices has measure zero, these are not strong restrictions in practice. In reverse mode AD, what we need is the adjoint relation $\overline{A} = \overline{A}(\overline{\lambda}, \overline{\vec{l}}, \overline{\vec{r}})$ as discussed in Sec. \ref{sec: Introduction}. Below we will adopt two different approaches to this task, namely the adjoint method and the more ``traditional'' approach based on the full eigen-decomposition process, and explain the intimate relation between them.

\subsection{\label{sec: the adjoint method}The adjoint method}
The adjoint method \cite{johnson2012notes} is a general way of deriving backward pass of various computation processes. To demonstrate the basic ideas, consider a simple yet generic setting as follows. Let $\vec{\theta} = (\theta_1, \cdots, \theta_P)$ be a $P$-dimensional input vector of parameters to be differentiated, and the output $\vec{x} = (x_1, \cdots, x_M)^T$ is an $M$-dimensional column vector. $\vec{x}$ is implicitly dependent on $\vec{\theta}$ through $M$ (generally nonlinear) equations of the form $f_i(\vec{x}, \vec{\theta}) = 0$, where $i$ ranges from $1$ to $M$. 

To derive the adjoint relation in the framework of reverse mode AD, one needs to compute the following vector-Jacobian product:
\begin{equation}
    \overline{\theta_\mu} = \overline{\vec{x}}^T \frac{\partial \vec{x}}{\partial \theta_\mu}, \quad \forall \mu = 1, \cdots, P.
    \label{eq: adjointp}
\end{equation}
where the $M$-dimensional column vector $\frac{\partial \vec{x}}{\partial \theta_\mu}$ is determined by the set of equations
\begin{equation}
    \frac{\partial f}{\partial \theta_\mu} + 
    \frac{\partial f}{\partial \vec{x}} \frac{\partial \vec{x}}{\partial \theta_\mu} = 0.
    \label{eq: partialxpartialp}
\end{equation}
where
\begin{equation}
    \frac{\partial f}{\partial \theta_\mu} = 
    \begin{pmatrix}
	    \frac{\partial f_1}{\partial \theta_\mu} \\ \vdots \\
	    \frac{\partial f_M}{\partial \theta_\mu}
    \end{pmatrix}, \quad
    \frac{\partial f}{\partial \vec{x}} = 
    \begin{pmatrix}
	    \text{---} & \frac{\partial f_1}{\partial \vec{x}} & \text{---} \\
	    & \vdots \\
	    \text{---} & \frac{\partial f_M}{\partial \vec{x}} & \text{---}
    \end{pmatrix}.
\end{equation}
Assuming the matrix $\frac{\partial f}{\partial \vec{x}}$ is invertible, one can solve for $\frac{\partial \vec{x}}{\partial \theta_\mu}$ directly from \Eq{eq: partialxpartialp} , then substitute it back to \Eq{eq: adjointp} to obtain the adjoint relation:
\begin{align}
	\overline{\theta_\mu} &= -\overline{\vec{x}}^T \left( \frac{\partial f}{\partial \vec{x}} 	\right)^{-1} \frac{\partial f}{\partial \theta_\mu} \nonumber \\
    				 &= -\vec{\eta}^T \frac{\partial f}{\partial \theta_\mu}.
	\quad \forall \mu = 1, \cdots, P.
	\label{eq: adjointp in general case}
\end{align}
where in the second line we have defined a column vector $\vec{\eta}$ determined by the so-called \emph{adjoint equation}:
\begin{equation}
    \left( \frac{\partial f}{\partial \vec{x}} 	\right)^T \vec{\eta} = \overline{\vec{x}}.
    \label{eq: general adjoint equation}
\end{equation}
In this way, we have also rearranged the order of matrix multiplication to avoid explicitly solving $\frac{\partial \vec{x}}{\partial \theta_\mu}$ appearing in \Eq{eq: adjointp}. This rearrangement is the core idea of the adjoint method. 

Specifically, in the settings of dominant eigen-decomposition described above, the $N$-dimensional matrix $A = A(\vec{\theta})$ depends on some parameters $\vec{\theta}$, and the output vector is effectively $(2N + 1)$-dimensional, including both the left/right eigenvectors $\vec{l}, \vec{r}$ and the scalar eigenvalue $\lambda$. The $2N + 1$ equations $f_i(\vec{l}, \vec{r}, \lambda, \vec{\theta}) = 0$ connecting the inputs and outputs are given by
\begin{equation}
    f_i(\vec{l}, \vec{r}, \lambda, \vec{\theta}) = 
    \begin{cases}
        (A - \lambda I)_i^T \vec{l}. & i = 1, \cdots, N. \\
        (A^T - \lambda I)_i^T \vec{r}. & i = N+1, \cdots, 2N. \\
        \vec{l}^T \vec{r} - 1. & i = 0.
    \end{cases}
    \label{eq: fs}
\end{equation}
where the subscript $i$ in an expression $M_i$ denotes the $i$th column of the matrix $M$, and the equation $f_0(\vec{l}, \vec{r}, \lambda, \vec{\theta}) = 0$ imposes the normalization constraint. 

Making use of \Eq{eq: fs}, one can solve for $\vec{\eta}$ in \Eq{eq: general adjoint equation}, then substitute it back to \Eq{eq: adjointp in general case} to obtain the desired expression of $\overline{\theta_\mu}$ for the dominant eigensolver. The derivation is fairly straightforward, and we refer the reader to Appendix \ref{appendix: adjoint method} for details. The final results are:
\begin{equation}
    \overline{\theta_\mu} = \overline{\lambda} \vec{l}^T \frac{\partial A}{\partial \theta_\mu} \vec{r}
                       -\vec{l}^T \frac{\partial A}{\partial \theta_\mu} \vec{\xi}_{\vec{l}}
                       -\vec{\xi}_{\vec{r}}^T \frac{\partial A}{\partial \theta_\mu} \vec{r}.
    \label{eq: adjointp for dominant eigen-decomposition}
\end{equation}
where the vectors $\vec{\xi}_{\vec{l}}$ and $\vec{\xi}_{\vec{r}}$ satisfy the linear systems
\begin{subequations}
    \begin{gather}
        (A - \lambda I) \vec{\xi}_{\vec{l}} = (1 - \vec{r}\vec{l}^T) \overline{\vec{l}}, \quad \vec{l}^T \vec{\xi}_{\vec{l}} = 0.
        \label{eq: lambda_0_l} \\
        (A^T - \lambda I) \vec{\xi}_{\vec{r}} = (1 - \vec{l}\vec{r}^T) \overline{\vec{r}}, \quad \vec{r}^T \vec{\xi}_{\vec{r}} = 0.
        \label{eq: lambda_0_r}
    \end{gather}
    \label{eq: lambda_0_l and lambda_0_r}
\end{subequations}
respectively. These linear systems are \emph{low-rank} in the sense that the coefficient matrices $A - \lambda I$ and $A^T - \lambda I$ are singular. Specifically, under our assumption that the eigenvalue $\lambda$ is non-degenerate, they have rank $N - 1$. Nevertheless, the solution for $\vec{\xi}_{\vec{l}}$($\vec{\xi}_{\vec{r}}$) is unique, because the singular matrix $A - \lambda I$($A^T - \lambda I$), when represented in the $(N - 1)$-dimensional subspace spanned by the $N - 1$ right(left) eigenvectors other than $\vec{r}$($\vec{l}$), is effectively non-singular. See also the discussions in Appendix \ref{appendix: adjoint method}.

\Eq{eq: adjointp for dominant eigen-decomposition} can be further simplified. In fact, we can ``strip'' the parameter $\vec{\theta}$ out of the primitive and obtain the neater expression for $\overline{A}$ by taking account of the fact that $\overline{\theta_\mu} = \textrm{Tr}\left(\overline{A}^T \frac{\partial A}{\partial \theta_\mu}\right)$. This way, we finally write the adjoint relation of dominant eigensolver as follows:
\begin{equation}
    \overline{A} = \overline{\lambda} \vec{l} \vec{r}^T - \vec{l} \vec{\xi}_{\vec{l}}^T - \vec{\xi}_{\vec{r}} \vec{r}^T.
    \label{eq: adjoint of general dominant eigen-decomposition}
\end{equation}
Fairly simple.

From Eqs. (\ref{eq: lambda_0_l and lambda_0_r}) and (\ref{eq: adjoint of general dominant eigen-decomposition}), one can see that the adjoint of $A$ needs only the desired eigenvalue $\lambda$ and corresponding eigenvectors $\vec{l}$, $\vec{r}$ without explicit reference to the full spectrum. In other words, we have successfully stripped out the information we want in the backward pass of dominant eigensolver, at the price of solving two somewhat nontrivial low-rank linear systems shown in \Eq{eq: lambda_0_l and lambda_0_r}. In a typical implementation, the forward pass can be accomplished by using Lanczos or other dominant eigen-decomposition algorithms, while the low-rank linear systems (\ref{eq: lambda_0_l and lambda_0_r}) involved in the backward pass can be solved efficiently using Krylov-based iterative algorithms such as biconjugate gradient and generalized minimal residual methods, among others. It's worth noting that both the dominant eigensolvers in the forward pass and the iterative linear system solvers in the backward pass do not need to know each individual entries of the matrix $A$; they only require the computation of matrix-vector products $A\vec{v}$ with an arbitrary vector $\vec{v}$. In many applications, this computation can be fairly efficient with the help of certain inner structures of $A$, even though the size of $A$ can be quite large.

\subsection{\label{sec: the special case where A is symmetric} Special case: $A$ is symmetric}
In this section, we will briefly discuss the special and important case where the real matrix $A$ to be diagonalized is symmetric. This case is particularly relevant to applications in quantum physics. There, all physical observables, including the Hamiltonian, are represented by Hermitian operators, thus also symmetric when all the matrix elements involved are real.

When $A$ is real symmetric, the desired left eigenvector is equal to the corresponding right eigenvector, that is, $\vec{l} = \vec{r} \equiv \vec{v}$. To obtain the adjoint of $A$ as a function of $\overline{\lambda}$ and $\overline{\vec{v}}$ in this special case, one can imitate the derivation in Sec. \ref{sec: the adjoint method} based on the adjoint method and obtain
\begin{gather}
    \overline{A} = (\overline{\lambda} \vec{v} - \vec{\xi}) \vec{v}^T, \quad \textrm{where $\vec{\xi}$ satisfies} \nonumber \\
    (A - \lambda I) \vec{\xi} = (1 - \vec{v} \vec{v}^T) \overline{\vec{v}}, \quad \vec{v}^T \vec{\xi} = 0.
    \label{eq: adjoint of symmetric dominant eigen-decomposition}
\end{gather}
This result can also be easily obtained from the general formulas (\ref{eq: lambda_0_l and lambda_0_r}) and (\ref{eq: adjoint of general dominant eigen-decomposition}). To do this, simply let the adjoint $\overline{\vec{l}}$ be zero, and $\overline{\vec{r}}$ be equal to $\overline{\vec{v}}$. The reason is that only the right eigenvector $\vec{r}$ is needed for downstream calculations, while the left eigenvector $\vec{l}$, which is equal to $\vec{r}$ in this case, acts as a piece of redundant information that doesn't affect the downstream results at all. It is then easy to see that the vector $\vec{\xi}_{\vec{l}}$ in \Eq{eq: lambda_0_l} vanishes, and the general formula (\ref{eq: adjoint of general dominant eigen-decomposition}) immediately reduces to the special form (\ref{eq: adjoint of symmetric dominant eigen-decomposition}).

It's instructive to furthermore inspect the physical implications of the adjoint relation (\ref{eq: adjoint of symmetric dominant eigen-decomposition}). Let again the matrix $A$ depend on one certain parameter, say, $\theta$. Then the adjoint of $\theta$ reads
\begin{align}
    \overline{\theta} &\equiv \textrm{Tr} \left( \overline{A}^T \frac{\partial A}{\partial \theta} \right) \nonumber \\
                 &= \overline{\lambda} \vec{v}^T \frac{\partial A}{\partial \theta} \vec{v} - \vec{\xi}^T \frac{\partial A}{\partial \theta} \vec{v}.
    \label{eq: adjointp in symmetric case}
\end{align}
Note the two terms correspond to dependence of the eigenvalue $\lambda$ and eigenvector $\vec{v}$ on $\theta$, respectively. In particular, if only the eigenvalue $\lambda$ is used for downstream computations, then the second term vanishes. For clarity, one could just consider the case where the loss $\mathcal{L} \equiv \lambda$. Thus we have $\overline{\lambda} = 1$, and \Eq{eq: adjointp in symmetric case} reduces to
\begin{equation}
    \overline{\theta} \equiv \frac{\partial \lambda}{\partial \theta} = \vec{v}^T \frac{\partial A}{\partial \theta} \vec{v}.
\end{equation}
This is the celebrated Hellmann-Feynman theorem \cite{PhysRev.56.340}, which is equivalent to the result of first order energy correction in perturbation theory. 
However, in the general case where the eigenvector also has nontrivial effect on the computation process, the second term in \Eq{eq: adjointp in symmetric case} is nonzero, and the formulation presented above turns out to be very useful.

\subsection{\label{sec: relation with the full eigen-decomposition approach}Relation with the full eigen-decomposition approach}

\begin{figure}[h]
    \includegraphics[page=1, width=0.85\columnwidth]{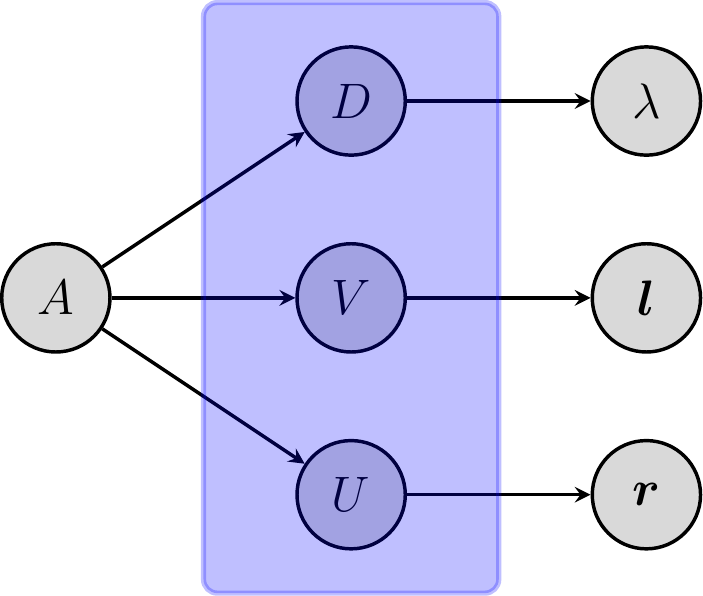}
    \caption{Studying the automatic differentiation of dominant eigensolver for a general real matrix $A$ by wrapping within it the corresponding full eigen-decomposition process. The internal data nodes $D$, $U$ and $V$, which are the outputs of the full eigen-decompostion process, act as a link for deriving desired results from already known ones. See text for more details, especially \Eq{eq: adjointD adjointU adjointalpha adjointx} for how this is achieved in reverse mode AD. It's worth noting that these inner structures are invisible if one treats the dominant eigensolver as a whole unit, which reflects the flexibility of the modular nature of differentiable programming paradigm.}
    \label{fig: wrap dominant eigen-decomposition}
\end{figure}

In Sec. \ref{sec: the adjoint method}, the automatic differentiation of dominant eigensolver has been presented straightforwardly in a full-spectrum-free form through the adjoint method. To figure out how this is achieved, it is instructive to change to another perspective by studying the relation between the adjoint method described above and the traditional approach based on full eigen-decomposition. The point is that we can wrap the process of full eigen-decomposition within the dominant one and utilize the results of the former formulation in the latter, as illustrated in \Fig{fig: wrap dominant eigen-decomposition}.

For clarity and without loss of generality, let $\lambda$ and $\vec{l}$, $\vec{r}$ be the ``first'' eigenvalue and eigenvectors of the $N$-dimensional matrix $A$, respectively. Recalling the assumption that $A$ is diagonalizable, we can write $V^T A U = D$, where
\begin{gather}
    D = \begin{pmatrix}
		\lambda \\
		& \lambda_2 \\
		& & \ddots \\
		& & & \lambda_N
	\end{pmatrix},
    U = \begin{pmatrix}
		\vert & \vert & & \vert \\
		\vec{r} & \vec{r}_2 & \cdots & \vec{r}_N \\
		\vert & \vert & & \vert
	\end{pmatrix}, \nonumber \\
	V^T \equiv U^{-1} = \begin{pmatrix}
	                        \text{---} & \vec{l}^T & \text{---} \\
	                        \text{---} & \vec{l}_2^T & \text{---} \\
	                        & \vdots \\
	                        \text{---} & \vec{l}_N^T & \text{---}
	                  \end{pmatrix}.
\end{gather}
That is, the columns of $U$ and rows of $U^{-1}$ correspond to the basis consisting of the $N$ right eigenvectors $(\vec{r}, \vec{r}_2, \cdots, \vec{r}_N)$ and left eigenvectors $(\vec{l}, \vec{l}_2, \cdots, \vec{l}_N)$, respectively.

In the framework of reverse mode AD, the adjoint relation of the full eigen-decomposition process is pretty standard \cite{giles2008extended} and reads
\begin{equation}
    \overline{A} = V \left[\overline{D} \circ I + (U^T \overline{U} - \overline{V}^T V) \circ F \right] U^T.
    \label{eq: full diagonalization AD}
\end{equation}
where $F$ is an anti-symmetric matrix with off-diagonal elements $F_{ij} = (\lambda_j - \lambda_i)^{-1}$ and $\circ$ denotes the Hadamard element-wise product. Here comes the key point. Since only $\lambda$, $\vec{l}$ and $\vec{r}$ will be used for downstream computations, the procedure of wrapping the process of full eigen-decomposition within the dominant one means that the adjoints of $D$, $U$ and $V$ should take the following form:
\begin{gather}
    \overline{D} \circ I = 
	    \begin{pmatrix}
		    \overline{\lambda} \\
		    & 0 \\
		    & & \ddots \\
		    & & & 0
	    \end{pmatrix}, 
    \overline{U} = 
	    \begin{pmatrix}
	    \begin{array}{c|}
		    \vert \\
		    \overline{\vec{r}} \\
		    \vert
	    \end{array} & 
        \begin{array}{ccc}
		    \\ & \Huge{0} & \\ & 
	    \end{array}
	    \end{pmatrix},
	\overline{V} = 
	    \begin{pmatrix}
	    \begin{array}{c|}
		    \vert \\
		    \overline{\vec{l}} \\
		    \vert
	    \end{array} & 
        \begin{array}{ccc}
		    \\ & \Huge{0} & \\ & 
	    \end{array}
	    \end{pmatrix}.
	\label{eq: adjointD adjointU adjointalpha adjointx}
\end{gather}
Substituting \Eq{eq: adjointD adjointU adjointalpha adjointx} into \Eq{eq: full diagonalization AD} yields
\begin{equation}
    \overline{A} = \overline{\lambda} \vec{l} \vec{r}^T - \sum_{i=2}^N c_i^{(\vec{r})} \vec{l}_i \vec{r}^T
                                                       - \sum_{i=2}^N c_i^{(\vec{l})} \vec{l} \vec{r}_i^T.
    \label{eq: adjoint of dominant eigen-decomposition another form}
\end{equation}
where we have introduced the quantities $c_i^{(\vec{r})} \equiv \frac{1}{\lambda_i - \lambda} \vec{r}_i^T \overline{\vec{r}}$ and $c_i^{(\vec{l})} \equiv \frac{1}{\lambda_i - \lambda} \vec{l}_i^T \overline{\vec{l}}, \forall i = 2, \cdots, N$. This formula looks quite similar to the earlier result (\ref{eq: adjoint of general dominant eigen-decomposition}) obtained by the adjoint method. Actually they are \emph{identically the same}. This can be seen by expanding the vectors $\vec{\xi}_{\vec{l}}$ and $\vec{\xi}_{\vec{r}}$ in \Eq{eq: lambda_0_l and lambda_0_r} in the complete basis $(\vec{r}, \vec{r}_2, \cdots, \vec{r}_N)$ and $(\vec{l}, \vec{l}_2, \cdots, \vec{l}_N)$, respectively. One can easily see that the quantities $c_i^{(\vec{l})}$ and $c_i^{(\vec{r})}$ defined above are exactly the linear combination coefficients of $\vec{\xi}_{\vec{l}}$ and $\vec{\xi}_{\vec{r}}$ in these two basis. In other words, we have
\begin{equation}
    \vec{\xi}_{\vec{l}} = \sum_{i=2}^N c_i^{(\vec{l})} \vec{r}_i, \quad
    \vec{\xi}_{\vec{r}} = \sum_{i=2}^N c_i^{(\vec{r})} \vec{l}_i.
\end{equation}
Plugging these relations back into \Eq{eq: adjoint of dominant eigen-decomposition another form} clearly reproduces the earlier result (\ref{eq: adjoint of general dominant eigen-decomposition}).

The observation above truely reveals the way in which the full spectrum information appearing explicitly in the original full eigen-decomposition approach can be effectively eliminated and replaced by the vectors $\vec{\xi}_{\vec{l}}$, $\vec{\xi}_{\vec{r}}$ characterized in \Eq{eq: lambda_0_l and lambda_0_r}. The fact that the final results of the two approaches are identically the same is not surprising, but the ``native'' representations are indeed different from a practical point of view. Specifically, the formulation based on the adjoint method clearly reveals the feasibility of constructing a valid dominant eigensolver primitive without any access to the full spectrum, while the approach based on full eigen-decomposition helps to furthermore clarify how this is achieved behind the scene.

\subsection{\label{sec: towards higher order derivatives}Towards higher order derivatives}

In this section, we study the possibility of performing higher order derivatives of the dominant eigensolver primitive. To do this, we have to investigate the backward pass of the low-rank linear system solvers described in \Eq{eq: lambda_0_l and lambda_0_r}, since this is the only non-trivial part in the backward pass of the primitive. It's instructive to study the simpler full-rank case first, where the coefficient matrix is non-singular. Specifically, Let $\vec{x}$ be the unique solution to the full-rank linear system $A \vec{x} = \vec{b}$, where $A$ is a non-singular matrix and $\vec{b}$ is an arbitrarily chosen vector. Since $A^{-1}$ exists, the derivation of the backward pass (i.e., the adjoint relation) is fairly straightforward. The final results are~\cite{giles2008extended}:
\begin{subequations}
    \begin{gather}
        \textrm{$\overline{\vec{b}}$ satisfies } A \overline{\vec{b}} = \overline{\vec{x}}, \\
        \overline{A} = - \overline{\vec{b}} \vec{x}^T.
    \end{gather}
    \label{eq: backward pass of full-rank linear system solver}
\end{subequations}
One can see that the backward pass of full-rank linear system solver involves solving another full-rank linear system. This observation is insightful, and as we will see, the similar conclusion can be drawn for the low-rank case.

The derivation for the backward pass of low-rank linear system solver is more subtle. For current purposes, it suffices to consider the following settings. Let $A$ be an $N$-dimensional real (diagonalizable) matrix of rank $N - 1$. This indicates that $A$ has $N - 1$ (right) eigenvectors $\vec{v}_2, \cdots, \vec{v}_N$ of nonzero eigenvalues $\lambda_2, \cdots, \lambda_N$, respectively, other than a single (right) eigenvector $\vec{v}$ with eigenvalue zero. For simplicity, we will restrict ourselves to the case where $A$ is symmetric, hence the left and right eigenvectors coincide. The derivation for the general case is pretty similar. Letting $\vec{b}$ be an arbitrary vector lying in the $(N - 1)$-dimensional subspace spanned by $\vec{v}_2, \cdots, \vec{v}_N$, the goal of the computation process is the unique solution for $\vec{x}$ of the following equations:
\begin{equation}
    A \vec{x} = \vec{b}, \quad \vec{v}^T \vec{x} = 0.
    \label{eq: low-rank linear system problem}
\end{equation}
These settings can fit properly into the backward pass of the low-rank linear system appearing in, say, \Eq{eq: adjoint of symmetric dominant eigen-decomposition}, under the correspondence $A \rightarrow A - \lambda I, \vec{x} \rightarrow \vec{\xi}, \vec{b} \rightarrow (1 - \vec{v} \vec{v}^T) \overline{\vec{v}}, \vec{v} \rightarrow \vec{v}$.

Rigorously speaking, the information about the eigenvector $\vec{v}$ of eigenvalue zero is contained in the matrix $A$. However this information is somewhat hard to extract directly, and in practice one finds it more convenient to treat $\vec{v}$ as an independent input to the process. To derive the adjoint relations for the low-rank linear system solver under these settings, one way is to manually perform decomposition of relevant quantities into the two orthogonal subspaces spanned by $\vec{v}$ alone and other $N-1$ eigenvectors of nonzero eigenvalues, respectively. This makes it more convenient to take advantage of the fact that $A$ is effectively invertible in the latter subspace. For more details, see Appendix \ref{appendix: low-rank linear system solver}. The final results are:
\begin{subequations}
    \begin{gather}
        \textrm{$\overline{\vec{b}}$ satisfies }A \overline{\vec{b}} = (1 - \vec{v} \vec{v}^T) \overline{\vec{x}}, \quad 
        \vec{v}^T \overline{\vec{b}} = 0.
        \label{eq: another low-rank linear system} \\
        \overline{A} = - \overline{\vec{b}} \vec{x}^T. \\
        \overline{\vec{v}} = - \vec{x} \vec{v}^T \overline{\vec{x}}.
    \end{gather}
    \label{eq: backward pass of low-rank linear system solver}
\end{subequations}

Notice the high similarity between \Eq{eq: backward pass of low-rank linear system solver} and the corresponding results (\ref{eq: backward pass of full-rank linear system solver}) for the full-rank case. Just as in the full-rank case, the backward pass of the low-rank linear system solver involves solving another low-rank linear system (\ref{eq: another low-rank linear system}) of the same kind. This observation is crucial and satisfying for the purposes of this paper. It implies that the backward pass of dominant eigensolver, which involves solving a low-rank linear system, can be conveniently differentiated itself by solving another low-rank linear system of the same kind, which in turn can be differentiated again, and so on. In other words, the formulation presented above allows us to compute in principle arbitrarily higher order derivatives of the dominant eigen-decomposition process in the framework of (reverse mode) automatic differentiation.

\section{\label{sec: Applications}Applications}
In this section, we demonstrate the use of dominant eigensolver primitive in action by two examples. Our code implementation \cite{github} is based on PyTorch \cite{pytorch}, a deep learning library that supports reverse mode AD through dynamic construction of computation graphs \cite{NIPS2019_9015}. In addition, note that PyTorch supports computing high order derivatives and, thanks to the modular nature of differentiable programming, has good flexibility and extensibility by allowing users to customize their own primitives. These features turns out to be very convenient for the purposes of this work.

Both of the examples are concentrated on the spin-$\frac{1}{2}$ transverse field Ising model (TFIM) on a $1$-dimensional lattice. The Hamiltonian reads:
\begin{equation}
    H = - \sum_{i=0}^{N-1} \left( g \sigma_i^x + \sigma_i^z \sigma_{i+1}^z \right).
    \label{eq: TFIM}
\end{equation}
where $g$ is a non-negative parameter characterizing the strength of the transverse magnetic field. When $g = 0$, the model reduces to the ``Ising limit'', and the spins in the ground state are perfectly aligned along the $z$ direction. When $g > 0$, on the other hand, the transverse field term will disrupt the magnetic order by introducing flipping of the spins. This model is a well-known prototype of the study of quantum phase transitions \cite{sachdev_2011}. Specifically, there is a transition point at $g = 1$ in thermodynamic limit, where the energy gap between ground state and the lowest excited state, which characterizes the energy scale of fluctuations at zero temperature for a gapped Hamiltonian like \Eq{eq: TFIM}, vanishes through a power law. Furthermore, this behavior can be characterized by a critical exponent, which usually turns out to be universal, that is, independent of most of the microscopic details of the system. 

Below we give a brief study of the model through the approach of exact diagonalization and variational optimization of matrix product states, respectively, using the formulations developed in Sec. \ref{sec: Formulations}.

\subsection{\label{sec: exact diagonalization} Identify the transition point by differentiating through exact diagonalization}
We first study the ground state properties of the model through exact diagonalization, specifically the behavior near the transition point. As indicated above, one of the main characterization of quantum phase transition (of gapped systems) is the vanishing of the gap between ground state and the lowest excited state in thermodynamic limit. There has been various kind of quantities proposed in practice to indicate the emergence of such behavior, and we have chosen two of them for the purpose of demonstration.

\subsubsection{\label{subsec: 2nd order derivative of the ground state energy}$2$nd order derivative of the ground state energy}
Computing the second order derivative of ground state energy (per site) $\frac{\partial^2 E_0}{\partial g^2}$ with respect to the parameter $g$ is a convenient way to characterize the quantum phase transition. In fact, the vanishing gap at the transition point implies the divergence of this quantity, which can be easily seen from the expression of $2$nd order perturbation theory as follows:
\begin{equation}
    \frac{\partial^2 E_0}{\partial g^2} = \sum_{n \neq 0} \frac{\left| \Braket{\psi_n | h^\prime | \psi_0}\right|^2}{E_0 - E_n}.
    \label{eq: d2E0}
\end{equation}
where
\begin{equation}
    h^\prime \equiv \frac{1}{N} \frac{\partial H(g)}{\partial g} = - \frac{1}{N} \sum_{i=0}^{N-1} \sigma_i^x.
\end{equation}
is the ``perturbation Hamiltonian (per site)'', and $H(g) \ket{\psi_n(g)} = N E_n(g) \ket{\psi_n(g)}$, with $n=0$ corresponding to the ground state.

Note that as discussed in Sec. \ref{sec: the special case where A is symmetric}, the computation of the $1$st order derivative of $E_0$ is more or less trivial and essentially equivalent to the Hellmann-Feynman theorem. Specifically, we have in current case
\begin{equation}
    \frac{\partial E_0}{\partial g} = \Braket{\psi_0 | h^\prime | \psi_0}.
    \label{eq: 1st derivative of E0}
\end{equation}
On the other hand, it is the computation of the $2$nd (and even higher) order derivative of $E_0$ that truly reveals the value of the machinery of automatic differentiation developed above. In fact, the $1$st order derivative of $E_0$ has explicit dependence on the eigenvector $\Ket{\psi_0}$ as shown in \Eq{eq: 1st derivative of E0}, and this implies that the $2$nd order derivative of $E_0$ has to be computed through direct differentiation onto $\Ket{\psi_0}$. This is exactly when the formulation based on AD could help to avoid the costly full eigen-decomposition as explicitly desired in \Eq{eq: d2E0}. See also the discussion in the last paragraph of Sec. \ref{sec: the special case where A is symmetric}.

\begin{figure}[h]
    \includegraphics[width=\columnwidth]{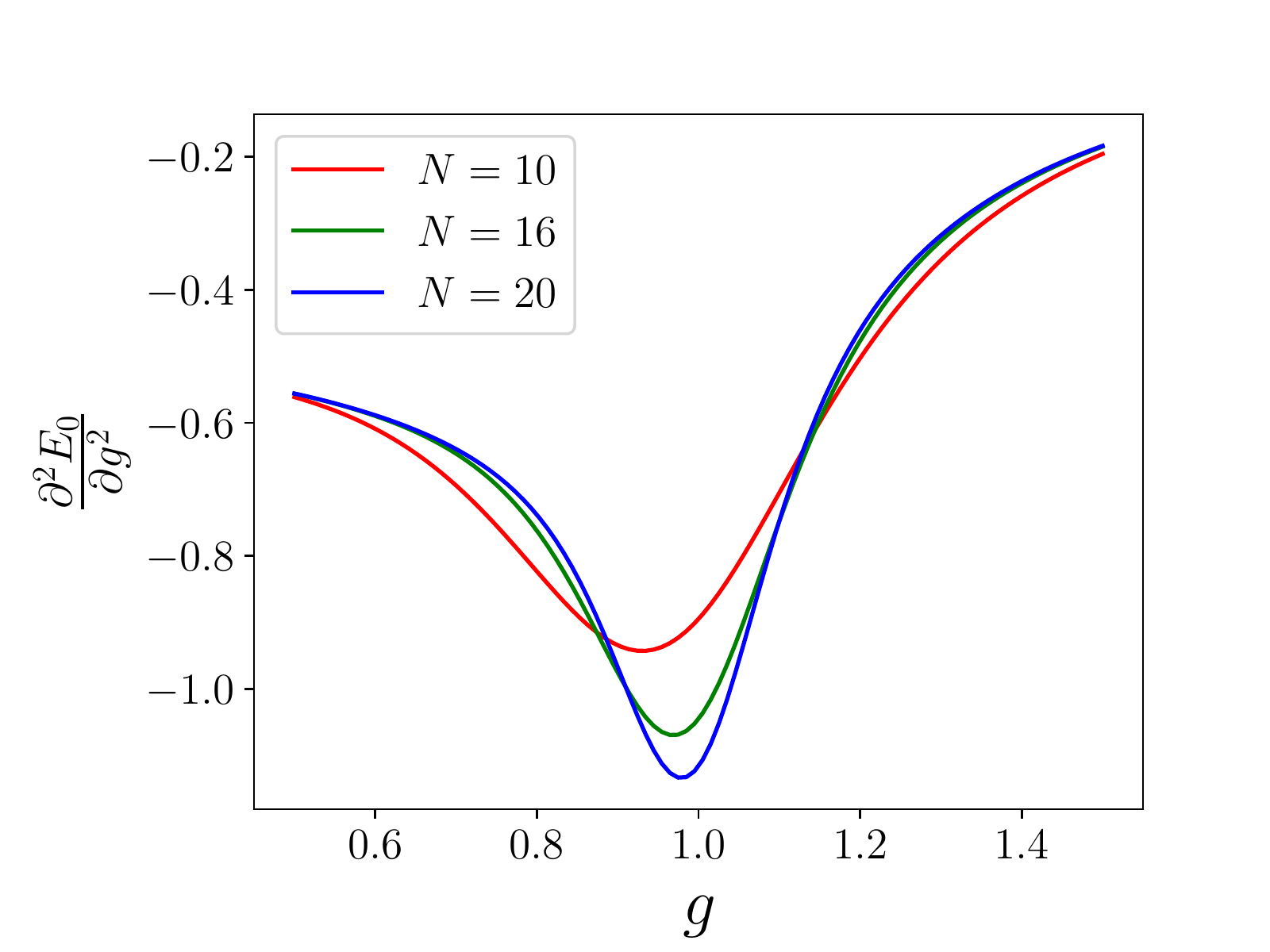}
    \caption{The $2$nd order derivative respect to parameter $g$ of the ground state energy per site $E_0$ of 1D TFIM for three values of the lattice size $N$, calculated through AD of the exact (dominant) digonalization.}
    \label{fig: d2E0}
\end{figure}

\Fig{fig: d2E0} shows $\frac{\partial^2 E_0}{\partial g^2}$ for three distinct lattice sizes $N$. Note that when $N = 20$,  the dimension of the Hilbert space involved is $2^{20} \sim 1000000$, and full eigen-decomposition of the Hamiltonian has become extremely challenging in practice \cite{PhysRevB.91.081103}. One can see that the $2$nd derivative of ground state energy is negative, which is a well-known fact in perturbation theory (See, e.g., \Eq{eq: d2E0}). In addition, the peak near the transition point $g = 1$ becomes more and more evident as $N$ increases, which agrees with the physical characterization of phase transition described above.

\subsubsection{\label{subsec: fidelity susceptibility}Fidelity susceptibility}
Another indicator of the quantum phase transition is the fidelity susceptibility \cite{PhysRevX.5.031007}, whose origin can be traced back to the field of quantum information science \cite{nielsen_chuang_2010}. To motivate this concept, note that there has emerged various kinds of classical and quantum phase transitions that go beyond the traditional Ginzburg-Landau-Wilson formulations based on the existence of some local order parameters. To name an example, topological phase transitions \cite{PhysRevLett.70.1501, PhysRevB.61.10267, KITAEV20062} do not have any local order parameter on either side of the phase transition. In view of this, new ideas and theoretical tools are needed to characterize these exotic phases and transitions among them, and various concepts in other fields, such as quantum fidelity \cite{doi:10.1142/S0217979210056335} and entanglement entropy \cite{RevModPhys.82.277} in quantum information science, have been borrowed and proved useful.

For current purposes, the concept of quantum fidelity is defined as the overlap of ground states of the Hamiltonian for two different parameters. More specifically, we have
\begin{equation}
    F(g, \varepsilon) = \left| \Braket{\psi_0(g) | \psi_0(g + \varepsilon)}\right|.
\end{equation}
Suppose the value of the parameter $g$ is fixed. When $\varepsilon = 0$, the two ground states coincide, and the quantum fidelity as a function of the ``distance'' $\varepsilon$ clearly reaches the maximum value $1$. Previous work \cite{PhysRevE.74.031123} has suggested that two ground states lying at different sides of a phase transition point is qualitatively different, thus have significantly smaller overlap. This means that when $g$ is near the transition point, the quantum fidelity as the function of $\varepsilon$ has more drastic changes at the maximum $\varepsilon=0$.

The concept of fidelity susceptibility $\chi_F$ is then proposed as a quantitative measure of this rate of change at $\varepsilon=0$ for various values of the parameter $g$. Specifically, it is defined as
\begin{equation}
    \chi_F = - \frac{\partial^2 \ln F(g, \varepsilon)}{\partial \varepsilon^2} \bigg{|}_{\varepsilon=0}.
    \label{eq: chiF original definition}
\end{equation}
It can be seen from the argument above that $\chi_F$ may exhibit a maximum or even diverge in thermodynamic limit at the transition point, which has been demonstrated by various works \cite{PhysRevLett.99.095701, Gu_2009, PhysRevLett.103.170501, PhysRevB.81.064418} and furthermore used for the detection and characterization of topological \cite{PhysRevA.78.010301, PhysRevA.79.032302, PhysRevB.80.014403} and other kinds of phase transitions.

Despite its high theoretical values, the practical calculation of fidelity susceptibility has become a difficult task in many situations, and many previous studies have thus been restricted to the case where the accurate ground state overlap can be computed via analytic results, exact diagonalization or density matrix renormalization group (DMRG) method. Even within the framework of exact diagonalization, the accurate computation of fidelity susceptibility is still annoying, which is largely due to the appearance of the $2$nd derivative in \Eq{eq: chiF original definition}. To make this statement clearer, one can do some simple manipulation on the original definition (\ref{eq: chiF original definition}) of $\chi_F$ and obtain an equivalent expression as follows:
\begin{equation}
    \chi_F = \Braket{\frac{\partial \psi_0}{\partial g} | \frac{\partial \psi_0}{\partial g}} - 
                \Braket{\psi_0 | \frac{\partial \psi_0}{\partial g}} \Braket{\frac{\partial \psi_0}{\partial g} | \psi_0}.
    \label{eq: chiF geometric form}
\end{equation}
where the ground state $\Ket{\psi_0(g)}$ has assumed to be normalized. The differential $\Ket{\frac{\partial \psi_0}{\partial g}}$ is difficult to handle properly, as already indicated by \Eq{eq: 1st order correction of ground state} and the discussions therein. The most natural approach is certainly through the perturbation theory, which yields
\begin{equation}
    \chi_F = \sum_{n \neq 0} \frac{\left| \Braket{\psi_n | h^\prime | \psi_0} \right|^2}{(E_0 - E_n)^2}.
    \label{eq: chiF perturbation form}
\end{equation}
Note that the second term in \Eq{eq: chiF geometric form} vanishes identically due to the orthogonality of $\Ket{\psi_0}$ and $\Ket{\frac{\partial \psi_0}{\partial g}}$ from the result of perturbation theory. \Eq{eq: chiF perturbation form} can be used as a benchmark against other methods based on certain approximations. However, it becomes intractable fairly quickly as the lattice size $N$ increases, due to the need of the full spectrum through highly expensive computation of full eigen-decomposition.

Comparing \Eq{eq: chiF perturbation form} with (\ref{eq: d2E0}) as well as the discussions therein, it is evident that the difficulties encountered here are pretty similar to those when attempting to compute the $2$nd derivative of the ground state energy in Sec. \ref{subsec: 2nd order derivative of the ground state energy}. Again, the formulation based on automatic differentiation provides a unifying and satisfactory solution. Specifically, one can directly implement the $2$nd order derivative in the original definition (\ref{eq: chiF original definition}) in the framework of reverse mode AD using PyTorch, by transforming it into a slightly different form:
\begin{equation}
    \chi_F = - \frac{\partial^2}{\partial g^{\prime 2}} \ln \left| \Braket{\psi_0(g) | \psi_0(g^\prime)} \right| \bigg{|}_{g^\prime = g}.
    \label{eq: chiF computation graph}
\end{equation}

\begin{figure}[h]
    \includegraphics[page=2]{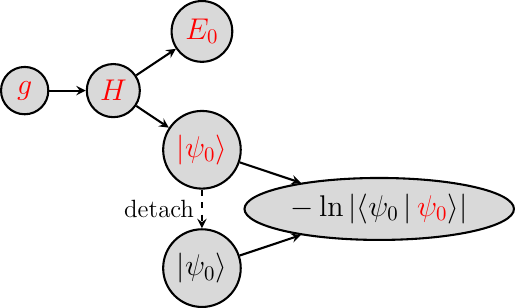}
    \caption{The graph of computing the fidelity susceptibility $\chi_F$ in the framework of reverse mode AD using \Eq{eq: chiF computation graph}. The color indicates the part of various data nodes that should be regarded as dependent on the input $g$ and differentiated upon when backwarding through the computation process. Note the presence of a detached duplicate of the ground state $\Ket{\psi_0}$, which is necessary to obtain the desired result.}
    \label{fig: chiF computation graph}
\end{figure}

\Fig{fig: chiF computation graph} shows the computation graph of this process. This graph has to be differentiated twice to obtain the fidelity susceptibility as indicated in \Eq{eq: chiF computation graph}. Note the partial derivative in (\ref{eq: chiF computation graph}) operates on only one of the two ground state vectors involved in the overlap. This implies that a \textit{detached} duplicate of $\Ket{\psi_0}$ has to be created, in the sense that the detached data node in a computation process is no longer treated as dependent on the inputs, as demonstrated by the dashed arrow in \Fig{fig: chiF computation graph}. This mechanism of detaching is well supported by PyTorch, which makes the implementation fairly straightforward and easy.

Since the output node appearing in \Fig{fig: chiF computation graph} has explicit dependence on the eigenvector $\Ket{\psi_0}$, the $1$st differentiation of the process involves solving low-rank linear systems of the kind in, say, \Eq{eq: adjoint of symmetric dominant eigen-decomposition}. This in turn makes the $2$nd differentiation involve the backward pass of this linear system solver, which typically involves the same kind of linear system solver again, as shown in Sec. \ref{sec: towards higher order derivatives}. To sum up, this example makes full use of the machinery developed in Sec. \ref{sec: Formulations}, including the mechanism of computing higher order derivatives.

\begin{figure}[h]
    \includegraphics[width=\columnwidth]{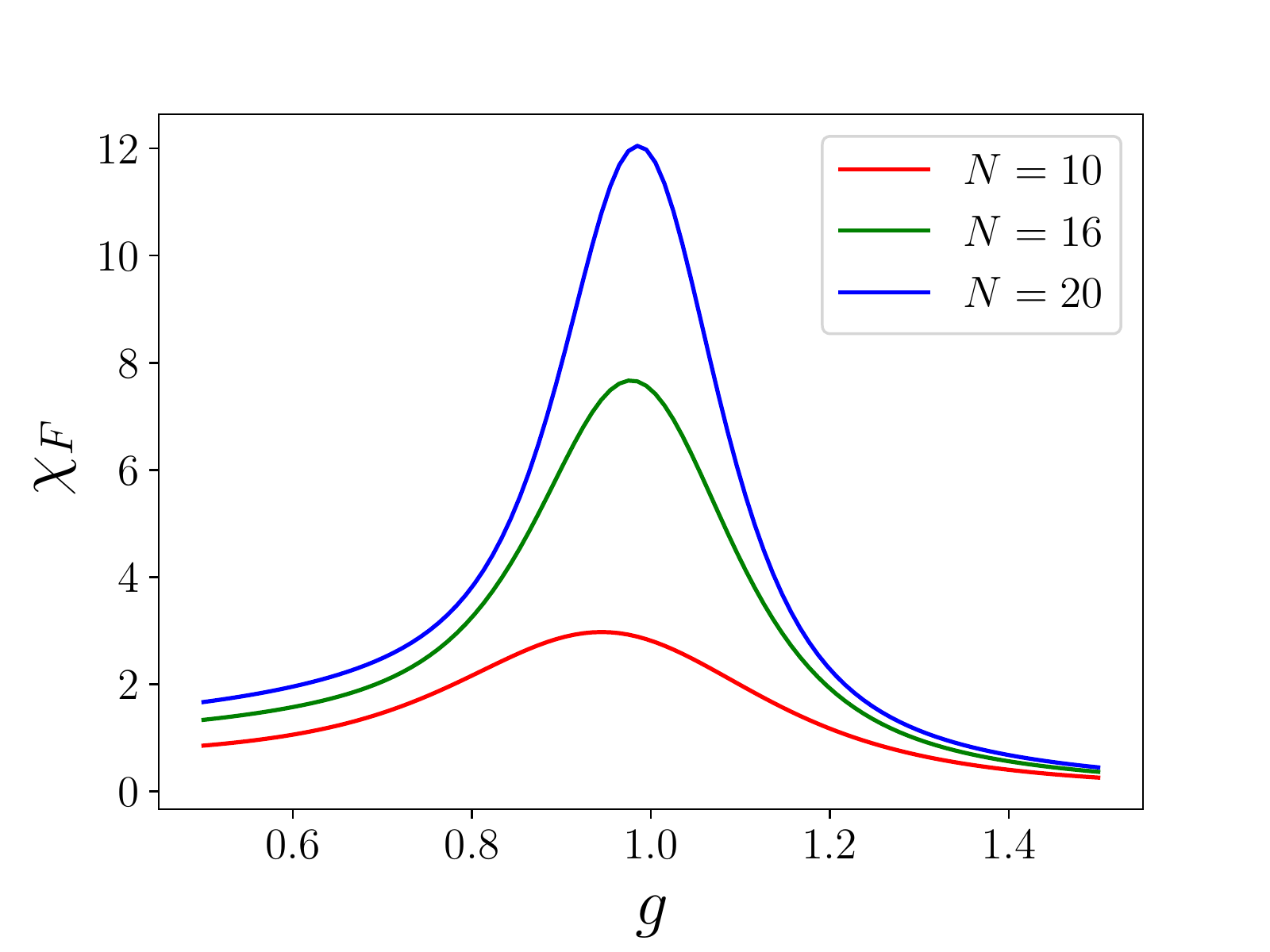}
    \caption{The fidelity susceptibility of 1D TFIM for three values of the lattice size $N$, calculated through AD of the exact (dominant) digonalization.}
    \label{fig: chiF}
\end{figure}
\Fig{fig: chiF} shows the computation results of $\chi_F$ for the same three distinct lattice sizes $N$ as in \Fig{fig: d2E0}. Observe that the fidelity susceptibility is always positive, which can be conveniently seen from \Eq{eq: chiF perturbation form}. In addition, there is indeed a peak near the transition point $g=1$ that grows higher quite rapidly as the lattice size increases. This agrees with the anticipation in previous studies regarding the good capability of fidelity susceptibility in detecting and characterizing various quantum phase transitions.

\subsection{\label{gradient optimization of uniform MPS}Gradient-based optimization of uniform MPS}
Matrix product states (MPS) originates from the celebrated DMRG method \cite{PhysRevLett.69.2863, PhysRevB.48.10345} and acts as the underlying variational ansatz of the formalism. MPS is a typical and well-known category of the rich family of tensor network (TN) states, which encode the correlation and entanglement of many-body states by virtual bonds connecting the microscopic degrees of freedom living on different sites. Based on such structures, TN states can provide a suitable parameterization of the low-energy states of various quantum many-body systems, with the theoretical guarantee relevant to the area law scaling of the entanglement entropy \cite{RevModPhys.82.277}. In particular, the class of MPS has proved to be very useful for studying the ground state of 1D strongly correlated systems with local interactions \cite{PhysRevB.73.094423, PhysRevB.76.035114}.

There have existed several schemes for the variational optimization of MPS states both for the finite and infinite, uniform lattice sites. Typical examples include various variations of the original DMRG algorithm \cite{SCHOLLWOCK201196}, infinite time evolving block decimation (iTEBD) based on Trotter decomposition of the evolution operator \cite{PhysRevLett.98.070201} and the recent variational uniform MPS (VUMPS) algorithm based on the concept of MPS tangent space \cite{PhysRevB.97.045145, 10.21468/SciPostPhysLectNotes.7}. Despite the maturity and successful applications of these methods in 1D systems, they usually lack generality and extensibility in some sense. For example, the formulation of VUMPS algorithm relies heavily on specific properties characteristic of the MPS state, and the analytic derivation of the gradients is rather cumbersome and error prone due to complicated and highly nonlinear dependence of the variational energy on input parameters. Maybe the most obvious consequence regarding this perspective is the well-known difficulty of generalization of these methods to two or higher dimensional systems.

The differentiable programming paradigm provides a natural solution in this respect. Owning to the inherent advantages mentioned in Sec. \ref{sec: Introduction}, differentiable programming serves as a suitable framework for various tensor network applications, and the practical implementation is usually more generic and free of specialized details relevant to certain settings. Recently \cite{PhysRevX.9.031041}, the technique of differentiable programming has been successfully used for studying ground state of the 2D square lattice antiferromagnetic Heisenberg model by gradient-base optimization of infinite projected entangled pair state (iPEPS). This clearly reveals its capability beyond other traditional tensor network methods.

Motivated by these arguments, we will calculate the ground state energy of 1D TFIM in thermodynamic limit by gradient-based variational optimization of uniform MPS, using the formulations developed in Sec. \ref{sec: Formulations}. The variational ansatz of the ground state reads
\begin{equation}
    \Ket{\psi_0} = \raisebox{-2.95ex}{\includegraphics[page=1]{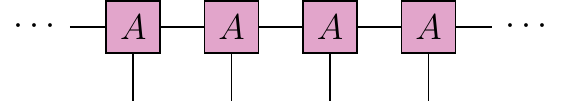}}.
    \label{eq: MPS ansatz}
\end{equation}
The parameter $A$ of the MPS is a rank-3 tensor of shape $d \times D \times D$, where $d = 2$ is the dimension of the local Hilbert space at each site, and $D$ is the virtual bond dimension. Note the model involves only nearest-neighbor interactions, which implies that the Hamiltonian (\ref{eq: TFIM}) can be written in the form
\begin{equation}
    H = \sum_{i=0}^{N-1} h_{i, i+1}.
\end{equation}
where $h_{i, j}$ is a ``local Hamiltonian'' that acts only on the two spin degrees of freedom at site $i$ and $j$. In the present case, for example, $h$ can be chosen to have the following symmetric form:
\begin{equation}
    h_{i, j} = -\frac{g}{2} \left( \sigma_i^x + \sigma_j^x \right) - \sigma_i^z \sigma_j^z.
\end{equation}

By simple algebraic manipulations, one can express the energy expectation value per site as follows:
\begin{equation}
    \frac{1}{N} \frac{\Braket{\psi_0 | H | \psi_0}}{\Braket{\psi_0 | \psi_0}} = \frac{1}{\lambda^2}
    \frac{\raisebox{0ex}{\includegraphics[page=2]{figs/MPS.pdf}}}
         {\raisebox{0ex}{\includegraphics[page=3]{figs/MPS.pdf}}}.
    \label{eq: MPS expectation energy per site}
\end{equation}
where $\raisebox{-5.5ex}{\includegraphics[page=4, scale=0.75]{figs/MPS.pdf}}
\equiv \Braket{s_i^\prime s_j^\prime | h_{i, j} | s_i s_j}$ is the tensor representation of $h_{i, j}$. $\lambda$ is the eigenvalue of largest amplitude of the transfer matrix $\raisebox{-3.2ex}{\includegraphics[page=5, scale=0.75]{figs/MPS.pdf}}$, and
$\raisebox{-2.5ex}{\includegraphics[page=6, scale=0.75]{figs/MPS.pdf}}$ and
$\raisebox{-2.5ex}{\includegraphics[page=7, scale=0.75]{figs/MPS.pdf}}$
are the corresponding left and right eigenvectors, respectively. More concretely, we have
\begin{equation}
    \raisebox{-7.5ex}{\includegraphics[page=8]{figs/MPS.pdf}} = \lambda
    \raisebox{-5.8ex}{\includegraphics[page=9]{figs/MPS.pdf}}, \quad 
    \raisebox{-7.5ex}{\includegraphics[page=10]{figs/MPS.pdf}} = \lambda
    \raisebox{-5.8ex}{\includegraphics[page=11]{figs/MPS.pdf}}.
    \label{eq: dominant eigen-decomposition of transfer matrix}
\end{equation}

Notice that the expectation energy \Eq{eq: MPS expectation energy per site} has nonlinear dependence on the parameter tensor $A$, both explicitly and implicitly via the eigenvalue $\lambda$ and eigenvectors $l$, $r$. We will compute the gradient of \Eq{eq: MPS expectation energy per site} with respect to $A$ using automatic differentiation, which can automatically take care of all the complicated ways of dependence without any laborious human efforts of deriving gradients analytically. 

Since the computation process involves the dominant eigen-decomposition (\ref{eq: dominant eigen-decomposition of transfer matrix}) of the transfer matrix and, notably, the output clearly has nontrivial dependence on the eigenvectors, the formulation developed in Sec. \ref{sec: Formulations} can thus be efficiently exploited. To put our contribution into the context of previous efforts~\cite{10.21468/SciPostPhysLectNotes.7}, it is instructive to manually back-propagate through the computation graph and derive the expression for the gradient with respect to $A$. For simplicity, one can safely set the normalization factor $\raisebox{-4ex}{\includegraphics[page=3, scale=0.75]{figs/MPS.pdf}}$ in \Eq{eq: MPS expectation energy per site} to $1$, which can always be achieved in practice. By making use of \Eq{eq: adjoint of general dominant eigen-decomposition}, one can then easily obtain
\begin{widetext}
    \begin{equation}
        \raisebox{-3ex}{\includegraphics[page=12]{figs/MPS.pdf}} \sim 
        \raisebox{-10.5ex}{\includegraphics[page=13]{figs/MPS.pdf}} + 
        \raisebox{-10.5ex}{\includegraphics[page=14]{figs/MPS.pdf}} + \overline{\lambda} \lambda^2 
        \raisebox{-10.5ex}{\includegraphics[page=15]{figs/MPS.pdf}} -
        \raisebox{-10.5ex}{\includegraphics[page=16]{figs/MPS.pdf}} -
        \raisebox{-10.5ex}{\includegraphics[page=17]{figs/MPS.pdf}}.
        \label{eq: MPS gradA expression}
    \end{equation}
\end{widetext}
where the first two and remaining terms correspond to the explict and implicit dependence on $A$, respectively. The vectors $\xi_l$, $\xi_r$ satisfy the kind of low-rank linear systems shown in \Eq{eq: lambda_0_l and lambda_0_r}, which in the present context have the forms
\begin{widetext}
    \begin{subequations}
        \begin{gather}
            \left( \raisebox{-7.15ex}{\includegraphics[page=18]{figs/MPS.pdf}} - \lambda \mathbbm{1}\right) 
            \raisebox{-5.5ex}{\includegraphics[page=19]{figs/MPS.pdf}} = 
            \left(\mathbbm{1} - \raisebox{-5.5ex}{\includegraphics[page=9]{figs/MPS.pdf}} \raisebox{-5.5ex}{\includegraphics[page=11]{figs/MPS.pdf}}\right) 
            \raisebox{-7.15ex}{\includegraphics[page=20]{figs/MPS.pdf}}, \quad
            \raisebox{-5.5ex}{\includegraphics[page=21]{figs/MPS.pdf}} = 0. \\
            \raisebox{-5.5ex}{\includegraphics[page=22]{figs/MPS.pdf}} 
            \left( \raisebox{-7.15ex}{\includegraphics[page=18]{figs/MPS.pdf}} - \lambda \mathbbm{1}\right) = 
            \raisebox{-7.15ex}{\includegraphics[page=23]{figs/MPS.pdf}}
            \left(\mathbbm{1} - \raisebox{-5.5ex}{\includegraphics[page=9]{figs/MPS.pdf}} \raisebox{-5.5ex}{\includegraphics[page=11]{figs/MPS.pdf}}\right), \quad 
            \raisebox{-5.5ex}{\includegraphics[page=24]{figs/MPS.pdf}} = 0.
        \end{gather}
        \label{eq: MPS low-rank linear systems}
    \end{subequations}
\end{widetext}

\Eq{eq: MPS gradA expression} is, apart from the gauge chosen, essentially equivalent to the gradient expression (116) in Ref.~\cite{10.21468/SciPostPhysLectNotes.7}. In particular, $\xi_l$, $\xi_r$ correspond to the quantities $R_h$ and $L_h$ in the reference, respectively. There, these two vectors are obtained by differentiating with respect to each tensor $A$ of the ansatz (\ref{eq: MPS ansatz}) and summing the resulting infinite series. The final expressions are given in Eq. (115) of the reference, which are essentially equivalent to the linear systems (\ref{eq: MPS low-rank linear systems}) above. Note that the similar pattern also appears when computing the single-site effective Hamiltonian in the VUMPS algorithm, which involves a step of formally performing some geometric sums. See \cite{PhysRevB.97.045145} for details. In the present work, on the other hand, the machinery of automatic differentiation through a dominant eigensolver allows us to bypass manually inspecting Eqs.~(\ref{eq: MPS gradA expression}, \ref{eq: MPS low-rank linear systems}) by encapsulating these derivations and calculations into a single computational primitive.  

\begin{figure}[h]
    \includegraphics[width=\columnwidth]{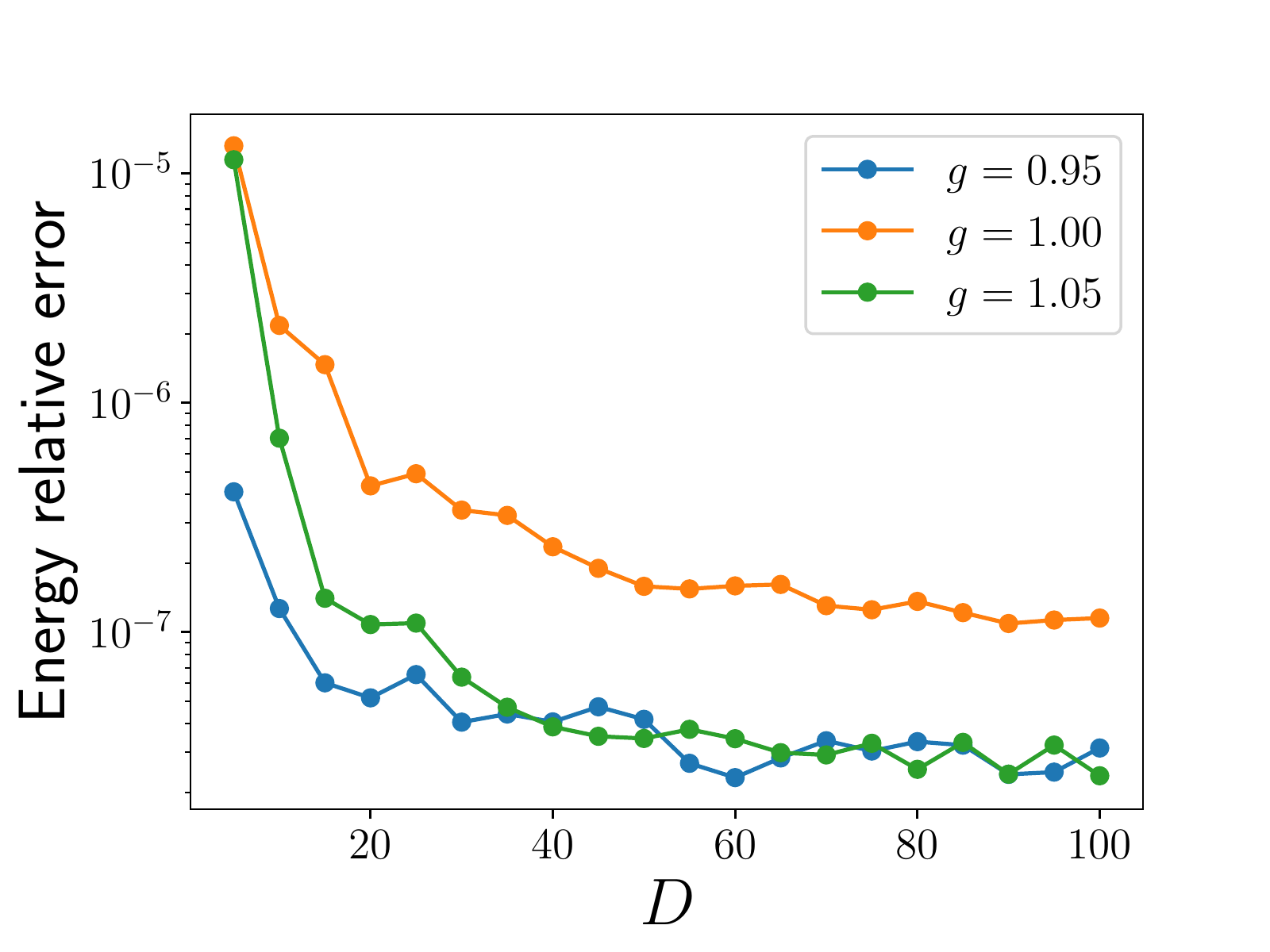}
    \caption{The relative error of the ground state energy per site $E_0$ of 1D TFIM for different values of the parameter $g$ and bond dimension $D$, compared to the analytic result \Eq{eq: E0 analytic result}.}
    \label{fig: E0 relative error}
\end{figure}

For current purposes, we don't impose any additional constraints on $A$ except assuming that it is real. Thus, the $D^2 \times D^2$ transfer matrix is generally not symmetric. Nevertheless, it is well-known that its largest-amplitude eigenvalue $\lambda$ is always real, positive and non-degenerate \cite{perez2006matrix}~\footnote{Strictly speaking, the largest-amplitude eigenvalue of the transfer matrix of an MPS is unique only if the MPS is injective. Nevertheless, since non-injective MPS tensors appear with measure zero, this is not a problem in practical computations. }, which meets the presupposition of the formulations in Sec. \ref{sec: Formulations}. In practice, the optimization is accomplished by using a quasi-Newton L-BFGS algorithm \cite{nocedal2006numerical} with automatically computed gradients. \Fig{fig: E0 relative error} shows the error of the ground state energy (per site) $E_0$ for several values of the parameter $g$ near the transition point $g=1$ and various bond dimensions $D$, relative to the analytic result obtained through Jordan-Wigner transformation \cite{sachdev_2011}:
\begin{equation}
    E_0(g) = - \frac{1}{2\pi} \int_0^{2\pi} \ud k \sqrt{g^2 - 2g\cos{k} + 1}.
    \label{eq: E0 analytic result}
\end{equation}
For each point in the figure, the convergence can be quickly reached after several hundreds of forward and backward pass, and the results are fairly accurate with relative errors of at most $10^{-5}$. Note the optimization is performed for bond dimension $D$ up to $100$, in which case the transfer matrix is of size $10000 \times 10000$ and the approach of full eigen-decomposition has become very slow. Another observation is that as one approaches the transition point $g=1$, it becomes more difficult to reach a certain level of accuracy. This phenomenon is typical and also arises in many other kinds of computational approaches, such as evaluating the integral \Eq{eq: E0 analytic result} and various quantum Monte Carlo methods.

However, \Fig{fig: E0 relative error} also has some drawbacks compared with the state-of-the-art VUMPS algorithm. In the present approach, the gradient (\ref{eq: MPS gradA expression}) is computed explicitly using automatic differentiation and then fed directly into a general-purpose optimizizer such as L-BFGS. Due to high complexity of the optimization landscape, it is generally very hard for such a optimizer to reach machine precision (i.e., $\sim 10^{-16}$) in a limited number of iteration steps. It's also easy to be trapped in local minima and, as a result, the converged energy is more or less sensitive to random initialization of $A$. In \Fig{fig: E0 relative error}, the strategy of initializing large $D$ simulation with the result of small one has been adopted to alleviate this issue. Nevertheless, the dependence of $E_0$ on $D$ is still slightly non-monotonous, which is not the case in VUMPS. In fact, the VUMPS algorithm is a more sophisticated variational approach by exploiting specific properties of the MPS manifold~\cite{PhysRevB.97.045145, 10.21468/SciPostPhysLectNotes.7} rather than simply employing a general-purpose optimization algorithm as we did here. Nevertheless, the approach based on automatic differentiation is clearly more straightforward and generic, and the results shown here are still satisfactory for the purpose of demonstration.

\section{\label{sec: Discussions}Discussions}

In this paper, the (reverse mode) automatic differentiation of dominant eigensolver is illustrated through two different yet equivalent approaches, namely the adjoint method and one based on the full eigen-decomposition process. In particular, the mechanism that effectively strips the desired information out of the full spectrum is carefully explained. In this respect, the former approach yields full-spectrum-free formulas more directly, while the results of the latter approach still have explicit dependence on the full spectrum. On the other hand, the latter approach reveals the modular nature of differentiable programming paradigm and is more routinely, while the former approach typically requires some specialized mathematical understanding of the primitive. In view of these arguments, the two approaches are complementary to each other and can be used for double checking in studying AD of certain computation processes.

In Sec. \ref{sec: Formulations}, we have taken into account only one eigenvalue and corresponding eigenvectors. The results presented there can be very easily generalized to the case of multiple eigenvalues and eigenvectors~\footnote{Actually, this is only true under the assumption that the desired eigenvalues are all non-degenerate. In the general case possibly with degeneracies, the formulas for AD of eigen-decomposition become considerably more complicated, and it's not straightforward to obtain a natural generalization of the present approach, particularly the low-rank linear systems shown in \Eq{eq: lambda_0_l and lambda_0_r}. However, it's worth noting that one may work around this issue for certain applications where the degenerate eigenvectors are used in downstream computations in such a way that the general AD formulas can be greatly simplified. In this case, a natural generalization of the formalism presented in this work is indeed possible. See also Ref.~\cite{PhysRevX.9.031041} for some discussions of this problem. }. In fact, different eigenvalues and eigenvectors are determined by the matrix $A$ in a totally independent way, implying that the adjoint of $A$ can be obtained simply by adding the contributions from them together, each of which has the form shown in Eqs. (\ref{eq: lambda_0_l and lambda_0_r}) and (\ref{eq: adjoint of general dominant eigen-decomposition}). The case of multiple eigenvalues and eigenvectors can be useful in many problems, such as Hamiltonian engineering to reproduce given low-energy spectrum \cite{PhysRevB.97.075114} and tensor network applications \cite{PhysRevX.9.031041}.

Furthermore, the formulations presented in this work can be readily extended to other similar computation processes. One typical and important example is the \emph{truncated} SVD, in which, as opposed to \emph{full} SVD, only a small proportion of singular values and corresponding singular vectors are desired. In practice, truncated SVD is widely adopted in various tensor network calculations \cite{ORUS2014117, SCHOLLWOCK201196}, such as the tensor renormalization group~\cite{PhysRevLett.99.120601} and corner transfer matrix renormalization group methods~\cite{doi:10.1143/JPSJ.65.891}. From the mathematical point of view, on the other hand, the truncated SVD of a real matrix $A$ has intimate relation with the dominant eigen-decomposition of the symmetric and positive semi-definite matrices $A A^T$ and $A^T A$. With this in mind, it turns out that one can exploit both of the approaches presented in Sec. \ref{sec: Formulations} in a similar way and derive the automatic differentiation of truncated SVD in a full-spectrum-free form. This will bring significant performance improvement over traditional approaches involving full SVD, such as the one implemented in Ref.~\cite{PhysRevX.9.031041}. Note that similar to the discussions in last paragraph, one can derive the relevant backpropagation formulas assuming only one singular value and corresponding singular vectors are desired. The generalization to the multiple case is then pretty straightforward.

\begin{acknowledgments}
The authors thank Shuo-Hui Li, Yue-Shui Zhang and Hai-Jun Liao for useful discussions. This work is supported by the National Natural Science Foundation of China under the Grant No.~11774398 and the Ministry of Science and Technology of China under the Grant No.~2016YFA0300603 and 2016YFA0302400.
\end{acknowledgments}

\appendix

\section{\label{appendix: adjoint method}Derivation of the backward pass of dominant eigensolver using the adjoint method}
In this appendix, we present derivation details of the backward pass of dominant eigensolver using the adjoint method introduced in Sec. \ref{sec: the adjoint method}. Under the settings of the dominant eigen-decomposition process described therein, one can make the correspondence with the generic notations appearing in Eqs. (\ref{eq: adjointp in general case}) and (\ref{eq: general adjoint equation}) as follows:
    \begin{widetext}
        \begin{equation}
        \vec{x} \rightarrow \begin{pmatrix}
    						    \vec{l} \\ \vec{r} \\ \lambda
    					    \end{pmatrix},
        \vec{\eta} \rightarrow \begin{pmatrix}
    									    \vec{\eta}_{\vec{l}} \\ \vec{\eta}_{\vec{r}} \\ \eta_\lambda
    							\end{pmatrix}, 
        \frac{\partial f}{\partial \theta_\mu} \rightarrow
        \begin{pmatrix}
    	    \frac{\partial f_1}{\partial \theta_\mu} \\ \vdots \\
    	    \frac{\partial f_N}{\partial \theta_\mu} \\
    	    \frac{\partial f_{N+1}}{\partial \theta_\mu} \\ \vdots \\
    	    \frac{\partial f_{2N}}{\partial \theta_\mu} \\
    	    \frac{\partial f_0}{\partial \theta_\mu}
        \end{pmatrix} = 
        \begin{pmatrix}
    	    \frac{\partial A^T}{\partial \theta_\mu} \vec{l} \\
    	    \frac{\partial A}{\partial \theta_\mu} \vec{r} \\ 0
        \end{pmatrix},
        \left( \frac{\partial f}{\partial \vec{x}} 	\right)^T \rightarrow
        \left(\begin{array}{ccc|ccc|c}
        	\vert & & \vert & \vert & & \vert & \vert \\
        	\left( \frac{\partial f_1}{\partial \vec{l}} \right)^T & \cdots & 
        	\left( \frac{\partial f_N}{\partial \vec{l}} \right)^T & 
        	\left( \frac{\partial f_{N+1}}{\partial \vec{l}} \right)^T & \cdots & 
        	\left( \frac{\partial f_{2N}}{\partial \vec{l}} \right)^T & 
        	\left( \frac{\partial f_0}{\partial \vec{l}} \right)^T \\
        	\vert & & \vert & \vert & & \vert & \vert \\ \hline
        	\vert & & \vert & \vert & & \vert & \vert \\
        	\left( \frac{\partial f_1}{\partial \vec{r}} \right)^T & \cdots & 
        	\left( \frac{\partial f_N}{\partial \vec{r}} \right)^T & 
        	\left( \frac{\partial f_{N+1}}{\partial \vec{r}} \right)^T & \cdots & 
        	\left( \frac{\partial f_{2N}}{\partial \vec{r}} \right)^T & 
        	\left( \frac{\partial f_0}{\partial \vec{r}} \right)^T \\
        	\vert & & \vert & \vert & & \vert & \vert \\ \hline
        	\frac{\partial f_1}{\partial \lambda} & \cdots & 
        	\frac{\partial f_N}{\partial \lambda} & 
        	\frac{\partial f_{N+1}}{\partial \lambda} & \cdots & 
        	\frac{\partial f_{2N}}{\partial \lambda} & 
        	\frac{\partial f_0}{\partial \lambda}
            \end{array}\right) = 
        \begin{pmatrix}
        	A - \lambda I & \vec{0} & \vec{r} \\
        	\vec{0} & A^T - \lambda I & \vec{l} \\
        	-\vec{l}^T & -\vec{r}^T & 0
        \end{pmatrix}.
        \label{eq: partial f partial theta correspondence}
        \end{equation}
    \end{widetext}
where the concrete form (\ref{eq: fs}) of the various equations $f_i(\vec{l}, \vec{r}, \lambda, \vec{\theta}) = 0$ has been used.   

The original adjoint equation (\ref{eq: general adjoint equation}) reduces to the following set of equations:
\begin{subequations}
    \begin{gather}
        (A - \lambda I) \vec{\eta}_{\vec{l}} + \eta_{\lambda} \vec{r} = \overline{\vec{l}}.
        \label{eq: lambda_l alone original} \\
        (A^T - \lambda I) \vec{\eta}_{\vec{r}} + \eta_{\lambda} \vec{l} = \overline{\vec{r}}.
        \label{eq: lambda_r alone original} \\
        -\vec{l}^T \vec{\eta}_{\vec{l}} - \vec{r}^T \vec{\eta}_{\vec{r}} = \overline{\lambda}.
        \label{eq: lambda_l lambda_r together}
    \end{gather}
    \label{eq: specific adjoint equation}
\end{subequations}
Taking account of the original relation (\ref{eq: relation between A, l, r}), one can first solve for $\eta_\lambda$ and obtains~\footnote{The equality between the last two quantities in \Eq{eq: lambda_alpha} is not so evident. However, we can reasonably accept this based on following arguments. Note that the condition (\ref{eq: relation between A, l, r}) doesn't uniquely determine the eigenvectors $\vec{l}$ and $\vec{r}$. In fact, it is invariant under the gauge transformation $\vec{l} \rightarrow \frac{\vec{l}}{c}, \vec{r} \rightarrow c \vec{r}$, where $c$ is an arbitrary non-zero scaling factor. This implies that any ``physical'' downstream computation must make use of $\vec{l}$ and $\vec{r}$ in a certain gauge-independent way, which will also impose some constraints on the adjoints $\overline{\vec{l}}$ and $\overline{\vec{r}}$. As a simple example, consider the case where the loss $\mathcal{L} = \vec{l}^T M \vec{r}$, where $M$ is an arbitrary square matrix. Then it's easy to check that $\vec{l}^T \overline{\vec{l}} = \vec{l}^T M \vec{r} = \vec{r}^T M^T \vec{l} = \vec{r}^T \overline{\vec{r}}$, as expected.}:
\begin{equation}
    \eta_\lambda= \vec{l}^T \overline{\vec{l}} = \vec{r}^T \overline{\vec{r}}.
    \label{eq: lambda_alpha}
\end{equation}
Substituting it back to the first two equations of \Eq{eq: specific adjoint equation} yields
\begin{gather*}
    (A - \lambda I) \vec{\eta}_{\vec{l}} = (1 - \vec{r} \vec{l}^T) \overline{\vec{l}}.
    \tag{\ref*{eq: lambda_l alone original}${}^\prime$} \label{eq: lambda_l alone} \\
    (A^T - \lambda I) \vec{\eta}_{\vec{r}} = (1 - \vec{l} \vec{r}^T) \overline{\vec{r}}.
    \tag{\ref*{eq: lambda_r alone original}${}^\prime$} \label{eq: lambda_r alone}
\end{gather*}
Note the solution for $\vec{\eta}_{\vec{l}}$ of \Eq{eq: lambda_l alone} alone is not unique. To make the solution structure clearer, we can separate out the component along $\vec{r}$ and write
\begin{equation}
    \vec{\eta}_{\vec{l}} = c_{\vec{l}} \vec{r} + \vec{\xi}_{\vec{l}}.
\end{equation}
Under the assumption that the matrix $A$ (hence $A - \lambda I$) is diagonalizable, the vector $\vec{\xi}_{\vec{l}}$ can be chosen to lie within the $(N-1)$-dimensional subspace spanned by the $N-1$ right eigenvectors other than the desired right eigenvector $\vec{r}$ of eigenvalue $\lambda$. This way, the value of $\vec{\xi}_{\vec{l}}$ becomes unique. Similar decomposition can be performed on the vector $\vec{\eta}_{\vec{r}}$ too. To summarize, one can obtain that
\begin{gather}
    \vec{\eta}_{\vec{l}} = c_{\vec{l}} \vec{r} + \vec{\xi}_{\vec{l}}, \quad \textrm{where $\vec{\xi}_{\vec{l}}$ satisfies} \nonumber \\
    (A - \lambda I) \vec{\xi}_{\vec{l}} = (1 - \vec{r}\vec{l}^T) \overline{\vec{l}}, \quad \vec{l}^T \vec{\xi}_{\vec{l}} = 0. \\
    \vec{\eta}_{\vec{r}} = c_{\vec{r}} \vec{l} + \vec{\xi}_{\vec{r}}, \quad \textrm{where $\vec{\xi}_{\vec{r}}$ satisfies} \nonumber \\
    (A^T - \lambda I) \vec{\xi}_{\vec{r}} = (1 - \vec{l}\vec{r}^T) \overline{\vec{r}}, \quad \vec{r}^T \vec{\xi}_{\vec{r}} = 0.
\end{gather}
These are the general solutions of \Eq{eq: lambda_l alone}, (\ref{eq: lambda_r alone}) for $\vec{\eta}_{\vec{l}}$ and $\vec{\eta}_{\vec{r}}$, respectively, where the coefficients $c_{\vec{l}}$ and $c_{\vec{r}}$ can take arbritrary values. However, under the additional condition \Eq{eq: lambda_l lambda_r together}, they must fulfill
\begin{equation}
    c_{\vec{l}} + c_{\vec{r}} = -\overline{\lambda}.
    \label{eq: cl and cr}
\end{equation}

As the final step, the generic equation (\ref{eq: adjointp in general case}) of the adjoint $\overline{\theta_\mu}$ of the parameters reduces to
\begin{align}
    \overline{\theta_\mu} &= -\vec{l}^T \frac{\partial A}{\partial \theta_\mu} \vec{\eta}_{\vec{l}}
                       -\vec{\eta}_{\vec{r}}^T \frac{\partial A}{\partial \theta_\mu} \vec{r} \nonumber \\
                          &= \overline{\lambda} \vec{l}^T \frac{\partial A}{\partial \theta_\mu} \vec{r}
                       -\vec{l}^T \frac{\partial A}{\partial \theta_\mu} \vec{\xi}_{\vec{l}}
                       -\vec{\xi}_{\vec{r}}^T \frac{\partial A}{\partial \theta_\mu} \vec{r}.
\end{align}
where we have used the correspondence (\ref{eq: partial f partial theta correspondence}) and \Eq{eq: cl and cr}. This is precisely the result shown in \Eq{eq: adjointp for dominant eigen-decomposition}.

\section{\label{appendix: low-rank linear system solver}Derivation of the backward pass of low-rank symmetric linear system solver}
In this appendix, we present derivation details of the backward pass of low-rank symmetric linear system solver (\ref{eq: low-rank linear system problem}) under the setting described therein. Recall that ($\vec{v}_2, \cdots, \vec{v}_N$) constitute the $N-1$ eigenvectors of the real symmetric matrix $A$ with nonzero eigenvalues $\lambda_2, \cdots, \lambda_N$, respectively, other than a single eigenvector $\vec{v}$ with eigenvalue zero. We introduce the following notations:
\begin{equation}
    D \equiv \begin{pmatrix}
		        \lambda_2 \\
		        & \ddots \\
		        & & \lambda_N
	        \end{pmatrix}, \quad 
    U \equiv \begin{pmatrix}
		        \vert & & \vert \\
		        \vec{v}_2 & \cdots & \vec{v}_N \\
		        \vert & & \vert
	        \end{pmatrix}.
\end{equation}
Note that $D$ is non-singular, since all of its diagonal elements are nonzero. It's not hard to see that they obey the following relations:
\begin{subequations}
    \begin{gather}
        A = U D U^T. \\
        U^T U = I_{N-1}. \\
        U U^T = I_N - \vec{v} \vec{v}^T. \label{eq: completeness relation}
    \end{gather}
    \label{eq: various relations}
\end{subequations}

From \Eq{eq: low-rank linear system problem}, we have
\begin{subequations}
    \begin{gather}
        A \ud \vec{x} = \ud \vec{b} - \ud A \vec{x}.
        \label{eq: derivative1} \\
        \ud \vec{v}^T \vec{x} + \vec{v}^T \ud \vec{x} = 0.
        \label{eq: derivative2}
    \end{gather}
\end{subequations}
Making use of the completeness relation (\ref{eq: completeness relation}), one could expand $\ud \vec{x}$ in the complete basis $(\vec{v}, \vec{v}_2, \cdots, \vec{v}_N)$ and get
\begin{equation}
    \ud \vec{x} = U U^T \ud \vec{x} + \vec{v} \vec{v}^T \ud \vec{x}.
\end{equation}
Basically, the conditions (\ref{eq: derivative1}) and (\ref{eq: derivative2}) completely determines the components of $\ud \vec{x}$ in the two orthogonal subspaces spanned by ($\vec{v}_2, \cdots, \vec{v}_N$) and $\vec{v}$ alone, respectively. Making use of the various relations in \Eq{eq: various relations}, it is easy to obtain
\begin{equation}
    \ud \vec{x} = U D^{-1} U^T (\ud \vec{b} - \ud A \vec{x}) - \vec{v} \vec{x}^T \ud \vec{v}.
\end{equation}
Comparing this result with the standard formula
\begin{equation}
    \ud \mathcal{L} = \overline{\vec{x}}^T \ud \vec{x} = 
    \mathrm{Tr}\left(\overline{A}^T \ud A\right) + \overline{\vec{b}}^T \ud \vec{b} + \overline{\vec{v}}^T \ud \vec{v}.
\end{equation}
one can immediately obtain the adjoint relations as follows:
\begin{subequations}
    \begin{gather}
        \overline{\vec{b}} = U D^{-1} U^T \overline{\vec{x}}.
        \label{eq: adjointb} \\
        \overline{A} = - U D^{-1} U^T \overline{\vec{x}} \vec{x}^T. \\
        \overline{\vec{v}} = - \vec{x} \vec{v}^T \overline{\vec{x}}.
    \end{gather}
\end{subequations}
Notice the explicit presence of the matrices $D$ and $U$, which contain the unknown and not desired information about the full spectrum of $A$. This is the price we have paid in order to conveniently make use of the fact that $A$ is effectively invertible in the subspace spanned by ($\vec{v}_2, \cdots, \vec{v}_N$), which is implied by the presence of $D^{-1}$ in the equations above.

To get rid of $D$ and $U$, note that the current situation is pretty similar to that we have encountered in Sec. \ref{sec: relation with the full eigen-decomposition approach} when deriving the backward pass of dominant eigensolver by using the results of full eigen-decomposition. Inspired by the discussions therein, we can just multiply both sides of \Eq{eq: adjointb} by $A$ and get
\begin{equation}
    A \overline{\vec{b}} = (1 - \vec{v} \vec{v}^T) \overline{\vec{x}}.
    \label{eq: linear system obeyed by adjointb}
\end{equation}
where we have used the relations (\ref{eq: various relations}) again. This equation is not completely equivalent to the original condition (\ref{eq: adjointb}) in the sense that it just characterizes the components of $\overline{\vec{b}}$ in the $(N-1)$-dimensional subspace spanned by ($\vec{v}_2, \cdots, \vec{v}_N$). To remedy this, just note that $\overline{\vec{b}}$ actually has zero component along the direction of the eigenvector $\vec{v}$, which can be seen from the right-hand side of (\ref{eq: adjointb}) directly. All that being said, we thus obtain the final adjoint relations for the low-rank symmetric linear system solver as follows, in a full-spectrum-free form:
\begin{subequations}
    \begin{gather}
        \textrm{$\overline{\vec{b}}$ satisfies }A \overline{\vec{b}} = (1 - \vec{v} \vec{v}^T) \overline{\vec{x}}, \quad 
        \vec{v}^T \overline{\vec{b}} = 0. \\
        \overline{A} = - \overline{\vec{b}} \vec{x}^T. \\
        \overline{\vec{v}} = - \vec{x} \vec{v}^T \overline{\vec{x}}.
    \end{gather}
\end{subequations}
This is precisely the results shown in \Eq{eq: backward pass of low-rank linear system solver}.


\bibliography{references}

\end{document}